\documentclass{desyproc}

\begin{document}
%------------------------------------
\title{Results and prospects on registration \\ of reflected Cherenkov light of EAS \\ from cosmic particles above $10^{15}$ eV}

%for single authors the superscripts are optional
\author{
{\slshape R.A. Antonov$^1$, T.V. Aulova$^1$,  E.A. Bonvech$^1$, D.V. Chernov$^1$, T.A. Dzhatdoev$^{1,*}$, \\ Mich. Finger$^{2,3}$, Mir. Finger$^{2,3}$, V.I. Galkin$^{4,1}$, D.A. Podgrudkov$^{4,1}$, T.M. Roganova $^1$}\\[1ex]
$^1$ Skobeltsyn Institute of Nuclear Physics, Lomonosov Moscow State University, Leninskie gory 1-2, 119991 Moscow, Russia \\
$^2$ Charles University in Prague, Faculty of Mathematics and Physics, V Holesovickach 2,
18000 Prague, Czech Republic\\
$^3$ Joint Institute for Nuclear Research, Joliot-Curie 6, 141980 Dubna, Moscow region, Russia  \\
$^4$ Faculty of Physics, Lomonosov Moscow State University, Leninskie gory 1-2, 119991 Moscow, Russia \\
$^*$ timur1606@gmail.com
}

% please enter the contribution ID for the DOI
\contribID{152}

% TO THE CONFERENCE EDITORS: 
% please update the following information      
% before sending the template to the authors
\confID{8648}  % if the conference is on Indico uncomment this line
\desyproc{DESY-PROC-2014-04}
\acronym{PANIC14} % if you want the Acronym in the page footer uncomment this line
\doi  % if there is an online version we will register DOIs

\maketitle

\begin{abstract}
We give an overview of the SPHERE experiment based on detection of reflected Vavilov-Cherenkov radiation ("Cherenkov light") from extensive air showers in the energy region E$>$10$^{15}$ eV. A brief history of the reflected Cherenkov light technique is given; the observations carried out with the SPHERE-2 detector are summarized; the methods of the experimental datasample analysis are described. The first results on the primary cosmic ray all-nuclei energy spectrum and mass composition are presented. Finally, the prospects of the SPHERE experiment and the reflected Cherenkov light technique are given. 
\end{abstract}

\section{Introduction \label{sec:intr}}

Despite several decades of intensive research, experimental results on the superhigh energy ($E>10^{15}$ eV = 1 PeV) cosmic ray spectrum and composition are still somewhat controversial. An uncertainty of the spectral shape is considerable (see, e.g., \cite{abb12}), and the results on the nuclear composition obtained by different experiments are often contradictory (e.g. \cite{tsu08}). A scatter of results is especially large for the composition studies: various measurements of the mean logarithmic mass number, $<ln A>$, at some energy region span almost the full range of masses from proton to Iron. For some extensive air shower (EAS) experimental techniques the uncertainty of the primary composition might translate into an additional error of the reconstructed all-nuclei spectrum, and this latter systematic uncertainty may dominate the total error of the spectrum measurement. 

The experimental situation clearly calls for development of new EAS observation and data analysis methods. In the present paper we describe one such method based on reflected Vavilov-Cherenkov radiation ("Cherenkov light") registration. Many details could be found in \cite{ant14}.

The first proposal to use a compact device lifted over a snow surface to observe reflected Cherenkov light of EAS was made by A.E. Chudakov \cite{chu74}. The first detector of such kind was developed by C. Castagnoli et al. \cite{cas81}. The first balloon-borne apparatus capable of reflected Cherenkov light observation was the SPHERE-1 detector \cite{ant97}; it had a mosaic of only 19 PMTs and could not register details of light impulse shape. The next generation, and currently the most advanced experiment with reflected Cherenkov light, employed the SPHERE-2 detector.

\section{The SPHERE-2 detector and the datasample \label{sec:dete}}

\begin{figure}[hb]
\centerline{\includegraphics[width=0.35\textwidth]{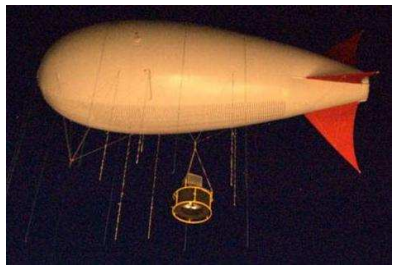}}
\caption{The SPHERE-2 detector carried by the BAPA tethered balloon.}\label{Fig:1}
\end{figure}

The SPHERE-2 balloon-borne detector \cite{ant14} had a mosaic of 109 PMTs and 12.5 ns time sampling (25 ns until the 2012 experimental run). A general view of the SPHERE-2 detector together with the BAPA tethered balloon is shown in Fig. 1. Observations were typically carried out in February-March at altitude H= 400--900 m above the surface of Lake Baikal. Total observation time for the 2008--2013 runs was about 140 h; about 1100 EAS were detected.

\section{Simulations, data analysis and results \label{sec:anal}}

\begin{figure}[hb]
\centerline{\includegraphics[width=0.30\textwidth]{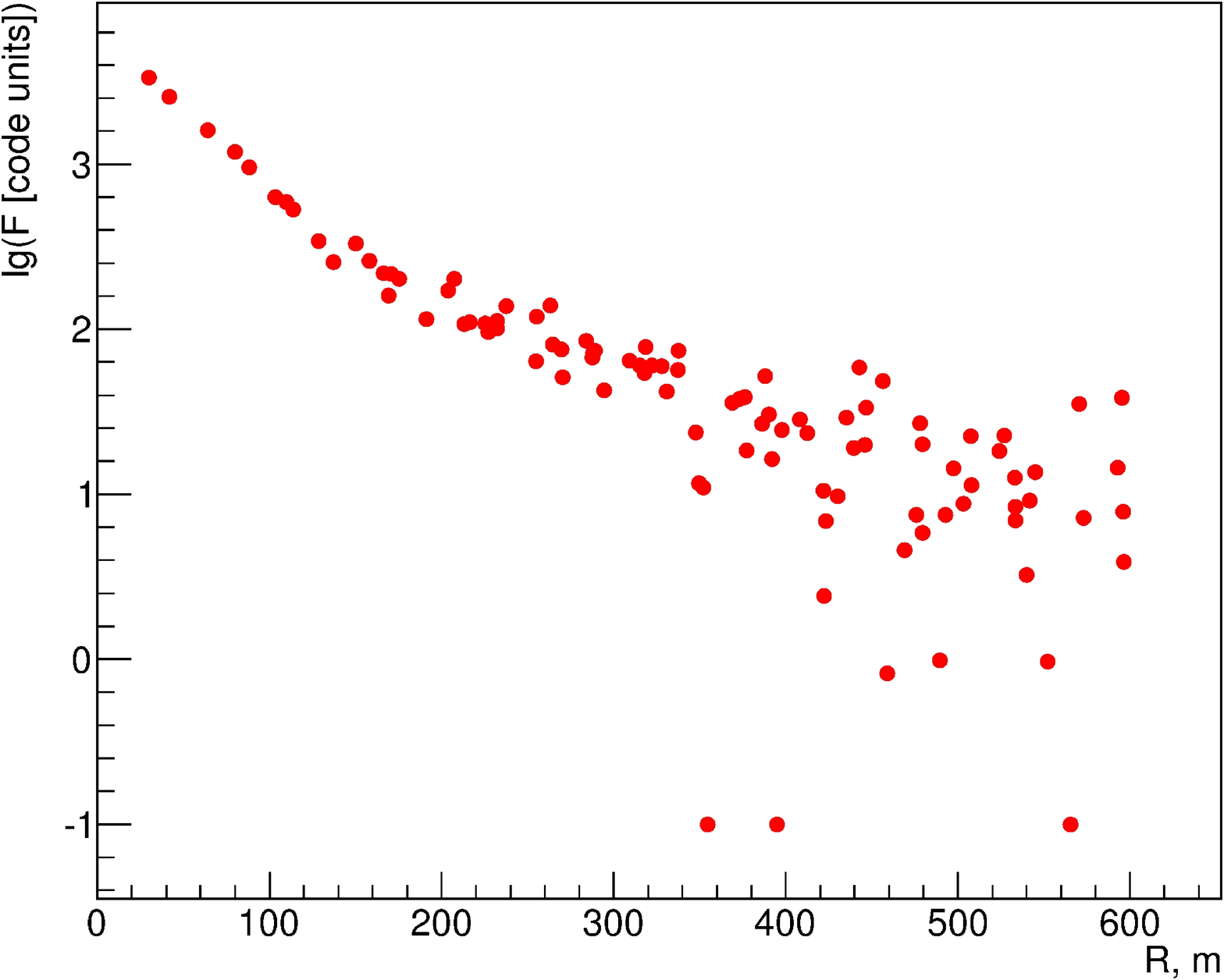} \includegraphics[width=0.40\textwidth]{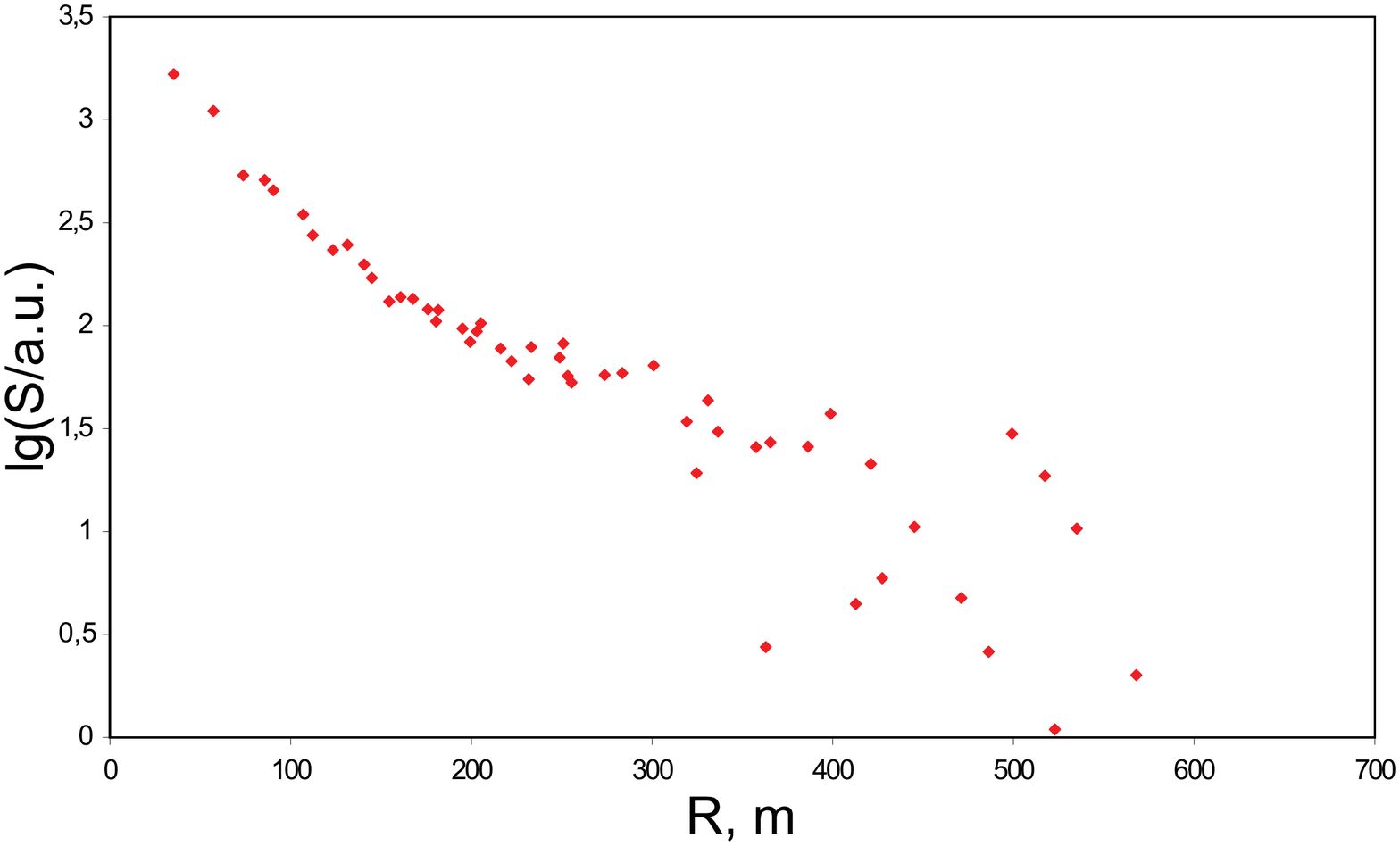}}
\caption{Two examples of reconstructed experimental LDFs: left --- for the first analysis, right --- for the second analysis.}\label{Fig:2}
\end{figure}

The detector response simulations were carried out by means of Monte Carlo (MC) approach using the CORSIKA code with the QGSJET-I high energy hadronic model and the GHEISHA low energy hadronic model \cite{hec98}, and the Geant4 code \cite{ago03}. The first step of experimental data analysis is reconstruction of lateral distribution function (LDF) of detected showers. An example of reconstructed LDF is shown in Fig. 2, left. LDF reconstruction is a quite complex procedure; it is important to have several independent techniques to ensure robustness of this procedure. An example of LDF obtained with the second technique is shown in Fig. 2, right. The next step of analysis, estimation of EAS primary energy, was performed by normalising the experimental LDFs to the model LDFs with known energy \cite{ded04}. Finally, the sample of the estimated primary energy values together with the model of the instrumental acceptance allowed the all-nuclei spectrum reconstruction \cite{ant13a},\cite{ant14}.

Simulated energy distributions for the 2013 experimental run data, power-law primary spectrum $J\sim E^{-\gamma}$ with slope $\gamma= 3$, and different types of primary nuclei, are shown in Fig. 3, left. The lowest curve corresponds to the primary Iron case, upper concentric curves are calculated for the Nitrogen, Helium and proton cases, correspondingly. The thick curve that fits the experimental histogram (circles) is drawn for the energy distribution with mixed composition. Information about the LDF steepness, that is sensitive to the primary composition, was utilized to build a model of energy distribution for mixed composition (see \cite{ant13a},\cite{ant14} for more details).

The combined all-nuclei spectrum for the 2011--2013 runs is shown in Fig.3 (right) by stars with statistical uncertainies (bars); systematic uncertainties are shown as well. For comparison the results of the KASCADE-Grande (triangles with associated statistical and systematic uncertainties) \cite{ape13a} and the Akeno (circles) \cite{nag92} experiments are shown. For the Akeno case statistical uncertainties are small and comparable to the diameter of the circles, and systematic uncertainties are unknown. For other results on the all-nuclei spectrum see \cite{aar13}.

The primary composition for the 2012 run was reconstructed using the LDF steepness parameter \cite{ant13b},\cite{ant14}, that allows an event-by-event composition study. The reconstructed composition for the 2012 run is shown in Fig. 2 of \cite{ant13b}. It is in general agreement with the KASCADE-Grande result \cite{ape13b}; the estimated fraction of light nuclei averaged over the 30--150 PeV energy region is 0.21$\pm$0.11.

\section{Prospects and conclusions \label{sec:conc}}

Uncertainty of results on the primary spectrum and composition discussed in sec. 3 at E$>$50 PeV is dominated by statistical errors. Two possible extensions of the SPHERE experiment to the energy region E$>$100 PeV were proposed \cite{ant14}: \\
1) A SPHERE-type detector with N$>$10$^3$ channels that would allow to perform an event-by-event study of the primary CR composition at E$>$100 PeV with statistical uncertainty comparable to the KASCADE-Grande's one given $\sim$500 h of exposition at H= 2--3 km; \\
2) A detector with N$>$10$^3$ channels aimed for study of the all-nuclei spectrum of Ultrahigh Energy Cosmic Rays (UHECR, E$>$10$^{18}$ eV) during a long-duration high-altitude (H$\approx$30--40 km) Antarctic flight, or, alternatively, a detector with N$>$10$^4$ channels that would allow to study the UHECR primary composition under similar experimental conditions.

\begin{figure}[t]
\centerline{\includegraphics[width=0.34\textwidth]{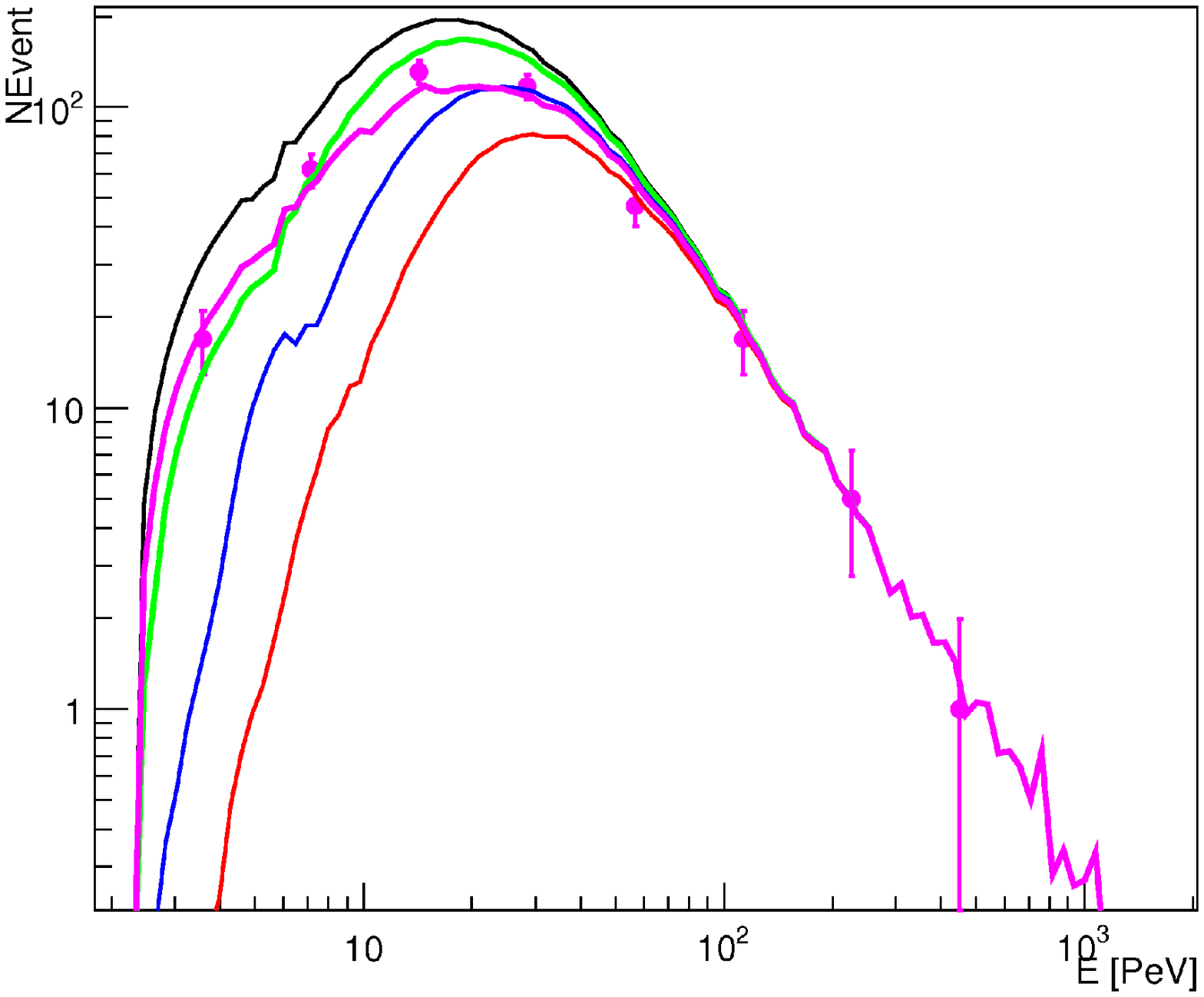} \includegraphics[width=0.35\textwidth]{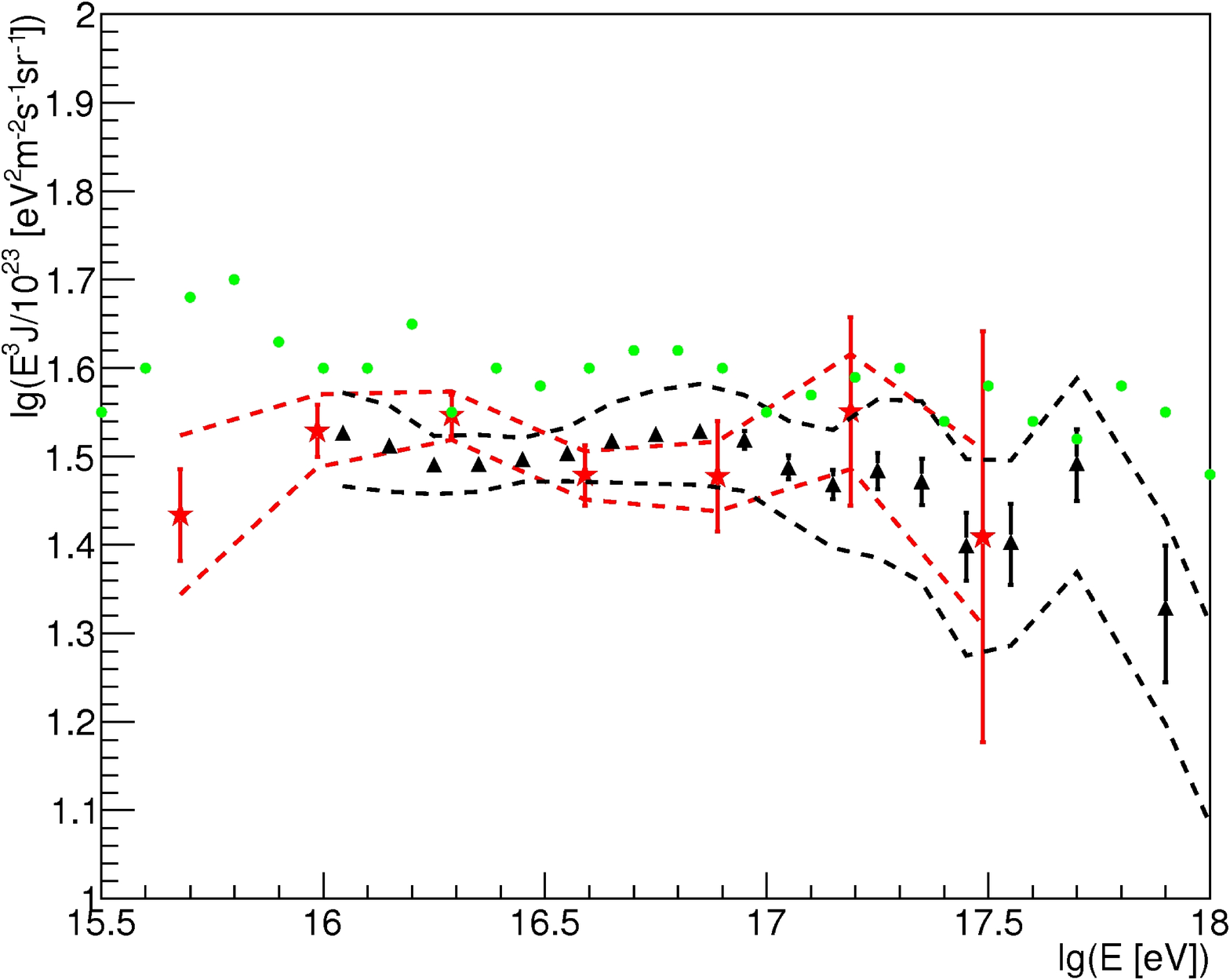}}
%{antonov_rem_fig4.eps}\includegraphics[width=0.35\textwidth]{antonov_rem_fig5.eps}}
\caption{Left --- model energy distribution for different composition assumptions (curves) and experimental energy distribution for the 2013 run data (circles). Right --- the all-nuclei spectrum reconstructed for the 2011--2013 runs (stars) along with the KASKADE-Grande and Akeno results.}\label{Fig:3}
\end{figure}

To conclude, we have reviewed a novel technique to study CR at E$>$1 PeV using reflected Cherenkov light. The method is currently mature enough to be competitive with other EAS observation methods, given sufficient observation time. For the first time, a detailed reconstruction of the all-particle CR spectrum at E= 3-300 PeV was performed using reflected Cherenkov light. This technique allows the CR nuclear composition study on event-by-event basis. Reflected Cherenkov light is a promising signal to study CR at E$>$100 PeV, either with tethered balloon at H= 2--3 km, or during a high-altitude Antarctic flight.

\section*{Acknowledgements}

The authors are grateful to the technical collaborators of the SPHERE-2 experiment. The work was supported by the Russian Foundation for Basic Research (grants 11-02-01475-a, 12-02-10015-k, 13-02-00470-а); the Russian President grants LSS-871.2012.2; LSS-3110.2014.2; the Program of basic researches of Presidium of Russian Academy of Sciences "Fundamental properties of a matter and astrophysics". Calculations were performed using the SINP MSU space monitoring data center computer cluster; we are grateful to Dr. V.V. Kalegaev for permission to use the hardware and to V.O. Barinova, M.D. Nguen, D.A. Parunakyan for technical support.
 
% ****************************************************************************
% BIBLIOGRAPHY AREA
% ****************************************************************************

\begin{footnotesize}
% IF YOU DO NOT USE BIBTEX, USE THE FOLLOWING SAMPLE SCHEME FOR THE REFERENCES
% ----------------------------------------------------------------------------

% ----------------------------------------------------------------------------

% IF YOU USE BIBTEX,
% - DELETE THE TEXT BETWEEN THE TWO ABOVE DASHED LINES
% - UNCOMMENT THE NEXT TWO LINES AND REPLACE 'Name_Of_Your_BibFile'

%\bibliographystyle{unsrt}
%\bibliography{Name_Of_Your_BibFile}
% example of Name_Of_Your_BibFile.bib
% @Article{Turcato:2006ch,
%      author    = "Turcato, M.",
%  collaboration = "ZEUS and H1",
%      title     = "Lepton flavour violation and charmonium physics at HERA",
%      journal   = "Nucl. Phys. Proc. Suppl.",
%      volume    = "162",
%      year      = "2006", 
%      pages     = "283-287",
%      SLACcitation  = "%%CITATION = NUPHZ,162,283;%%"
% }
% 
% @Unpublished{Gogitidze:2007du,
%      author    = "Gogitidze, N.",
%  collaboration = "H1", 
%      title     = "Prompt photons and particle momentum distributions at
%                   HERA", 
%      year      = "2007",
%      note    = "hep-ex/0701033",
%      SLACcitation  = "%%CITATION = HEP-EX 0701033;%%"
% }

\end{footnotesize}

% ****************************************************************************
% END OF BIBLIOGRAPHY AREA
% ****************************************************************************

\end{document}